\documentclass[twocolumn,preprintnumbers,amsmath,amssymb]{revtex4}
\usepackage{graphicx}
\usepackage{dcolumn}
\usepackage{bm}

\newcommand{\be}{\begin{equation}}
\newcommand{\bel}[1]{\begin{equation}\label{#1}}
\newcommand{\ee}{\end{equation}}
\newcommand{\bea}{\begin{eqnarray}}
\newcommand{\ba}{\begin{array}}
\newcommand{\eea}{\end{eqnarray}}
\newcommand{\ea}{\end{array}}

\begin{document}

\title{Asymmetric simple exclusion process describing conflicting traffic flows }

\author{M. Ebrahim Foulaadvand $^{1,2}$ and Mehdi Neek-Amal $^{2}$ }

\address{$^1$ Department of Physics, Zanjan University, P.O. Box
45196-313, Zanjan, Iran.\\
$^2$ Computational physical sciences research laboratory, Department of Nano-Sciences, Institute for studies in
theoretical Physics and Mathematics (IPM),\\
P.O. Box 19395-5531, Tehran, Iran.}

\begin{abstract}

We use the asymmetric simple exclusion process for describing
vehicular traffic flow at the intersection of two streets. No
traffic lights control the traffic flow. The approaching cars to
the intersection point yield to each other to avoid collision.
This yielding dynamics is model by implementing exclusion process
to the intersection point of the two streets. Closed boundary
condition is applied to the streets. We utilize both mean-field
approach and extensive simulations to find the model
characteristics. In particular, we obtain the fundamental
diagrams and show that the effect of interaction between chains
can be regarded as a dynamic impurity at the intersection point.

\end{abstract}

\pacs{PACS numbers: 05.60.-k, 05.50.+q, 05.40.-a, 64.60.-i }


\maketitle

\section{Introduction}

Modelling a vast variety of non equilibrium phenomena has
constituted the subject of intensive research by statistical
physicists \cite{zia,schutz1}. In particular, vehicular dynamics
has been one of these fascinating issues
\cite{kernerbook,css99,helbing}. While the existing results in
highway traffic needs further manipulations in order to find
direct applications, researches on {\it city traffic}
\cite{bml,nagatani,cs} seem to have more feasibility in practical
applications. Recently, notable attention have paid to
controlling traffic flow at intersections and other designations
such as roundabouts
\cite{foolad1,foolad2,foolad3,ray,xiong,gershenson,foolad4}. In
this respect, we intend to study another aspect of traffic flow at
intersections. In principle, the vehicular flow at an
intersection can be controlled via two schemes. In the first
scheme the traffic is controlled without traffic lights. In the
second scheme, signalized traffic lights control the flow. In the
former scheme, approaching car to the intersection yields to the
traffic in its perpendicular direction by adjusting its velocity
to a safe value to avoid collision. The basic question is that
under what circumstances the intersection should be controlled by
traffic lights? In order to capture the basic features of this
problem, we construct a simple stochastic model. The vehicular
dynamics is represented by {\it asymmetric simple exclusion
process }(ASEP) \cite{mcdonald,derrida1,derrida2}. The
intersection point is the place where two chains representing the
streets interact with each other. It is a well-established fact
that a single static impurity can strongly affect the
characteristics of ASEP both in closed \cite{lebowitz1,lebowitz2}
and open boundary condition \cite{kolomeisky1}. In addition, the
characteristics of ASEP in the presence of moving impurities has
been studied and shown to exhibit disorder-induced phase
transitions \cite{evans1}. Besides relevance to traffic flow, the
investigation of ASEP in the presence of small amount disorder
has recently revealed the existence of novel aspects of the
interplay of disorder and drive
\cite{chou,lakatos1,foolad5,pronina}. In our model, the effect of
the perpendicular chain can be interpreted as a single dynamic
site-wise disorder which to our knowledge has not been
investigated.

\section{ Description of the Problem }

Consider two perpendicular one dimensional closed chains each
having $L$ sites ($L$ is even). The chains represent urban roads
accommodating unidirectional vehicular traffic flow. They cross
each other at the crossing sites $i_1=i_2=\frac{L}{2}$ on the
first and the second chain respectively. With no loss of
generality we take the direction of traffic flow in the first
chain from south to north and in the second chain from east to
west (see Fig.1 for illustration). Each site of the chain is
either vacant or can hold at most one car (hereafter
interchangeably called particle). We assign an integer valued
occupation number $n_i$ ($m_i$) to site $i$ of the first (second)
chain respectively. In case the site is occupied by a particle,
its occupation number is one and zero otherwise.

\begin{figure}
\centering
\includegraphics[width=7.5cm]{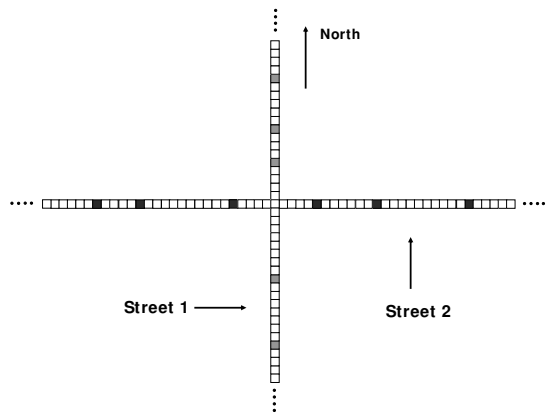}
\caption{Two perpendicular closed chains representing the
intersection of two uni directional traffic flows. } \label{fig:bz2}
\end{figure}

The system configuration at each time $t$ is characterized by
specifying the occupation numbers $n_i$ and $m_i$ ($i=1,\cdots,
L$) in both chains. The system dynamics is asymmetric simple
exclusion process (ASEP). Although this choice is far from the
realistic vehicular dynamics, we expect the general features of
the problem namely the phase structure does not qualitative
change. In addition, employing ASEP dynamics allows us to treat
the problem more easily on analytical grounds. In reality, each
driver yields to perpendicular traffic flow by appropriately
adjusting of her velocity. In our formulation, this cautionary
behaviour can be described by simple exclusion process. This
process is defined as follows. During an infinitesimal time $dt$
each particle can stochastically hop to its forward neighbouring
site provided the target site is empty. If the target site is
already occupied by another particle, the attempted movement is
rejected. At each time $t$, we define the particle approaching to
the intersection as the one occupying the previous site
$i=\frac{L}{2}-1$. We note that this particle's attempt is
successful if both sites $i_1=\frac{L}{2}$ and $i_2=\frac{L}{2}$
which is actually the common intersection site on both chains is
empty. In terms of ASEP terminology, the sites $\frac{L}{2}-1$
can be regarded as dynamic impurity sites for each chain. The
hoping rate at all sites are scaled to unity except at
$i_1=\frac{L}{2}-1$ and $i_2=\frac{L}{2}-1$ which are
$1-m_{\frac{L}{2}}$ and $1-n_{\frac{L}{2}}$ respectively.

\subsection{Mean field approach }

Let us denote the mean density at site $i$ and time $t$ of the
first and second chains by $\langle n_i \rangle$ and $\langle m_i \rangle$ respectively.
The master equation governing their time evolution can simply be written as
follows:\\

\be \frac{d}{dt} \langle n_i \rangle= \langle
n_{i-1}(1-n_i) \rangle - \langle n_{i}(1-n_{i+1}) \rangle ~~
\ee

In the above equation, the site index $i$ covers all the lattice
sites except $i=\frac{L}{2}-1$ and $i=\frac{L}{2}$. The rate
equation for the second chain can be obtained by replacing $n_i$
by $m_i$. For sites $i=\frac{L}{2}-1$ and $i=\frac{L}{2}$, the
rate equations have the following forms:

\be \frac{d}{dt} \langle n_{\frac{L}{2}-1} \rangle= \langle
n_{\frac{L}{2}-2}(1-n_{\frac{L}{2}-1}) \rangle - \langle
n_{\frac{L}{2}-1}(1-n_{\frac{L}{2}}-m_{\frac{L}{2}}) \rangle \ee

\be \frac{d}{dt} \langle n_{\frac{L}{2}} \rangle= \langle
n_{\frac{L}{2}-1}(1-n_{\frac{L}{2}}-m_{\frac{L}{2}}) \rangle -
\langle n_{\frac{L}{2}}(1-n_{\frac{L}{2}+1}) \rangle \ee

Similarly, the rate equations for $m_{\frac{L}{2}-1}$ and
$m_{\frac{L}{2}}$ can be obtained by replacing $m\leftrightarrow
n$ respectively. In order to proceed analytically we take into
account the mean-field approximation. In this approximation, we
replace the two-point functions by the product of one-point
functions and furthermore we replace the probability that the
middle site $i=\frac{L}{2}$ is empty, i.e.
$(1-n_{\frac{L}{2}}-m_{\frac{L}{2}})$, by
$(1-n_{\frac{L}{2}})(1-m_{\frac{L}{2}})$. The latter expression is
the probability that the middle sites in each chain are
simultaneously empty. In the steady state, the left hand sides of
mean-field equations become zero and we arrive at a set of $2L$
nonlinear equations. Even by employing the assumption of mean
field, we are not able to solve these nonlinear algebraic
equations. Therefore, we should resort to numerical methods. We
now outline a numerical approach for solving the set of nonlinear
equations.

\subsection{ Numerical approach to mean field equations }

Our approach for solving the MF equations is based on the
constant density scheme which has originally been introduced by
Barma and Tripathy \cite{barma1,barma2}. In this scheme, we first
fix the global densities in two chains at given values $\rho_1$
and $\rho_2$. Second, we assign initial density profiles
$n_1[0],n_2[0],\cdots,n_L[0]$ and $m_1[0],m_2[0],\cdots,m_L[0]$
to the first and second chain respectively. The constancy of
global densities implies the following constraints on the initial
profiles:

\be n_1[0]+n_2[0]+\cdots+n_L[0]=L\rho_1 \ee

\be m_1[0]+m_2[0]+\cdots+m_L[0]=L\rho_2 \ee

We next evolve the site densities according the following
discrete time updating rules:

\be n_i[t+1]=n_i[t]+ n_{i-1}[t](1-n_i[t]) - n_i[t](1-n_{i+1}[t])
\ee

Similar equations hold for $m_i$. Note that in the above
equations, $i$ covers the whole chain except the sites
$i=\frac{L}{2}-1$ and $i=\frac{L}{2}$. The above discrete time
evolution rules stem in time discretisation of the mean-field
equations within Euler algorithm. In the special sites
$i=\frac{L}{2}-1,\frac{L}{2}$ the interaction between two chains
modifies the rate equations and we should take into account
equations (2,3) which gives rise to the following dynamical rules:

$$ n_{\frac{L}{2}-1}[t+1]=n_{\frac{L}{2}-1}[t] +
n_{\frac{L}{2}-2}[t](1-n_{\frac{L}{2}-1}[t])-$$ \be
n_{\frac{L}{2}-1}[t](1-n_{\frac{L}{2}}[t])(1-m_{\frac{L}{2}}[t])
\ee

$$ n_{\frac{L}{2}}[t+1]=n_{\frac{L}{2}}[t] +
n_{\frac{L}{2}-1}[t](1-n_{\frac{L}{2}}[t])(1-m_{\frac{L}{2}}[t])-$$
\be n_{\frac{L}{2}}[t](1-n_{\frac{L}{2}+1}[t]) \ee

The equations for $m_{\frac{L}{2}-1}$ and $m_{\frac{L}{2}}$ are
simply obtained from the above equations via replacing $m$ by $n$
and vice versa. Notice that our dynamical equations preserve the
constancy of global densities. More concisely, we have
$n_1[t]+n_2[t]+\cdots+n_{L}[t]=L\rho_1$ and
$m_1[t]+m_2[t]+\cdots+m_{L}[t]=L\rho_2$ at each $t$.

With an appropriate choice of initial condition, after iterating
the above equations for many time steps the system is expected to
reach a fixed point denoted by $\{n_i^* \}$ and $\{ m_i^* \}$ in
which further iteration does not change the densities. This
solution can be considered as the solution of the mean-field
equations. Note that in a acceptable solution, all the densities $
n_1^*,\cdots,n_L^*$ and $m_1^*,\cdots,m_L^*$ should lie between
zero and one. The chains currents are thus obtained according to
the relations $J_1=n_i^*(1-n_{i+1}^*)$ and
$J_2=m_i^*(1-m_{i+1}^*)$ in which $i$ can be any site of the
chains. It should be noted that the choice of initial densities
is a crucial step. Only certain initial condition converges to
the desired solution. In general, the long-time behaviour of the
density profile turns out to be an oscillatory pattern.

\subsection{ Monte Carlo simulation}

For obtaining a better insight, we have also executed extensive
Monte Carlo simulations which are presented in this section. The
chains sizes are equally taken as $L_1=L_2=300$ and we averaged
over 100 independent runs each of which with $10^5$ time steps
per site. After transients, two chains maintain steady-state
currents denoted by $J_1$ and $J_2$ which are function of the
global densities $\rho_1$ and $\rho_2$. We kept the global
density at a fixed value $\rho_2$ in the second chain and varied
$\rho_1$. Figure (2) exhibits the fundamental diagram of the
first chain.

\begin{figure}
\centering
\includegraphics[width=7.5cm]{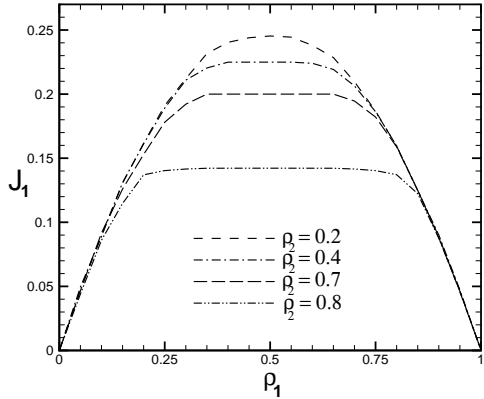}
\caption{First chain current $J_1$ versus its density $\rho_1$ for
various densities $\rho_2$ in the second chain. $L=300$. }
\label{fig:bz2}
\end{figure}

The generic behaviour is reminiscent to ASEP with a single
defective site \cite{lebowitz1,lebowitz2,barma2}. Intersection of
two chains makes the intersection point appear as a dynamical
defect. It is a well-known fact that a local defect can affect
the system on a global scale \cite{lebowitz2,kolomeisky1}. This
has been confirmed not only for simple exclusion process but also
for cellular automata models, such as Nagel-Schreckenberg
\cite{ns92}, describing vehicular traffic flow
\cite{css99,chung,yukawa}. Analogous to static defects, in our
case of dynamical impurity we observe that the effect of the
dynamic defect is to form a plateau region $ \rho \in [\rho_{-},
\rho_{+}]$ in which $\rho_{\pm}=0.5\pm\Delta$ and $2\Delta$ is
the extension of the plateau region in the fundamental diagram.
In the plateau region the current is independent of the global
density. The larger the density in the perpendicular chain is,
the dynamic defect has larger strength. For higher $\rho_2$, the
plateau region is wider and correspondingly the current value is
more reduced. The notable point is that the flow capacity in the
first chain persist to large decrease up to considerably large
density $\rho_2 \sim 0.5$ in the second chain. This marks the
fact that conflicting flows of particles can, to a large extent,
weakly affect each other. To shed some more light on this aspect,
let us now consider the flow characteristics in the second chain.
In figure (3) we sketch the behaviour of perpendicular fundamental
diagram that is $J_2$ versus $\rho_1$.

\begin{figure}
\centering
\includegraphics[width=7.5cm]{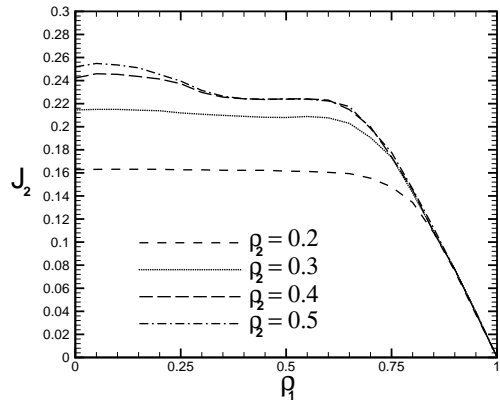}
\caption{Second chain current $J_2$ versus density $\rho_1$ of the
first chain for various densities $\rho_2$. } \label{fig:bz2}
\end{figure}

As depicted, $J_2$ shows a phase transition at the critical
density $\rho_{1,c}=\rho_{+}$. Before $\rho_{1,c}$, the current
exhibits smoothly decreasing behaviour. The nature of decrease
depends on the value of $\rho_2$. For $\rho_2 <0.3$ or $\rho_2
>0.7$ $J_2$ is almost constant and is obtained from the single
chain relation $J_2=\rho_2(1-\rho_2)$. However, in the interval
$0.3 < \rho_2 < 0.7~$ $J_2$ shows a complex behaviour as is shown
in figure (3). It first increases up to a small $\rho_2$ then
smoothly diminishes until it reaches to a plateau region. The
reason is due to interaction between two chains which induces
correlations between them. This modifies the value of $J_2$ from
the single chain value $\rho_2(1-\rho_2)$. The appearance of a
maximum in $J_2$ marks the point that a small density of cars in
the first chain can even regulate the traffic, and hence enhance
the flow, in the second chain. When the global density in the
first chain exceeds the critical value, the perpendicular current
exhibits a quasi linear decline. This corresponds to capacity
break down in the second chain. In terminology of vehicular
traffic, if the density of  the perpendicular chain goes beyond a
critical value, one should be warned that controlling of the
traffic via self-organised mechanism starts to fail and traffic
lights signalisation is thereby prescribed.

\subsection{density profiles}

In order to improve our understanding, it would be useful to look
at the behaviour of density profiles in both chains. Let us look
at some typical density profiles before attempting to give a
general remark. The following figures, obtained by simulation,
display the density profile in the first chain for two densities
$\rho_1=0.2$ and $\rho_1=0.6$.

\begin{figure}
\centering
\includegraphics[width=7.5cm]{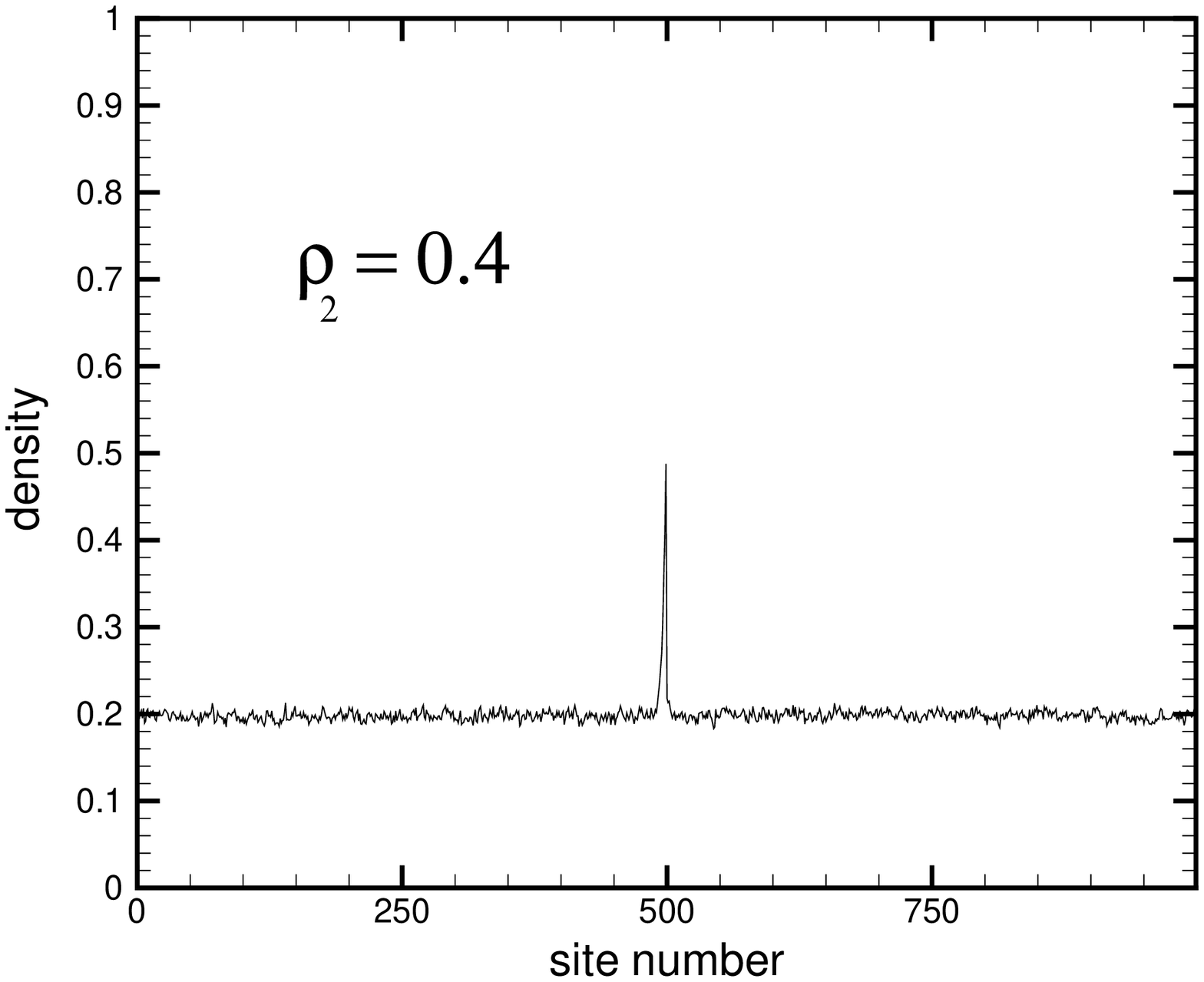}
\includegraphics[width=7.5cm]{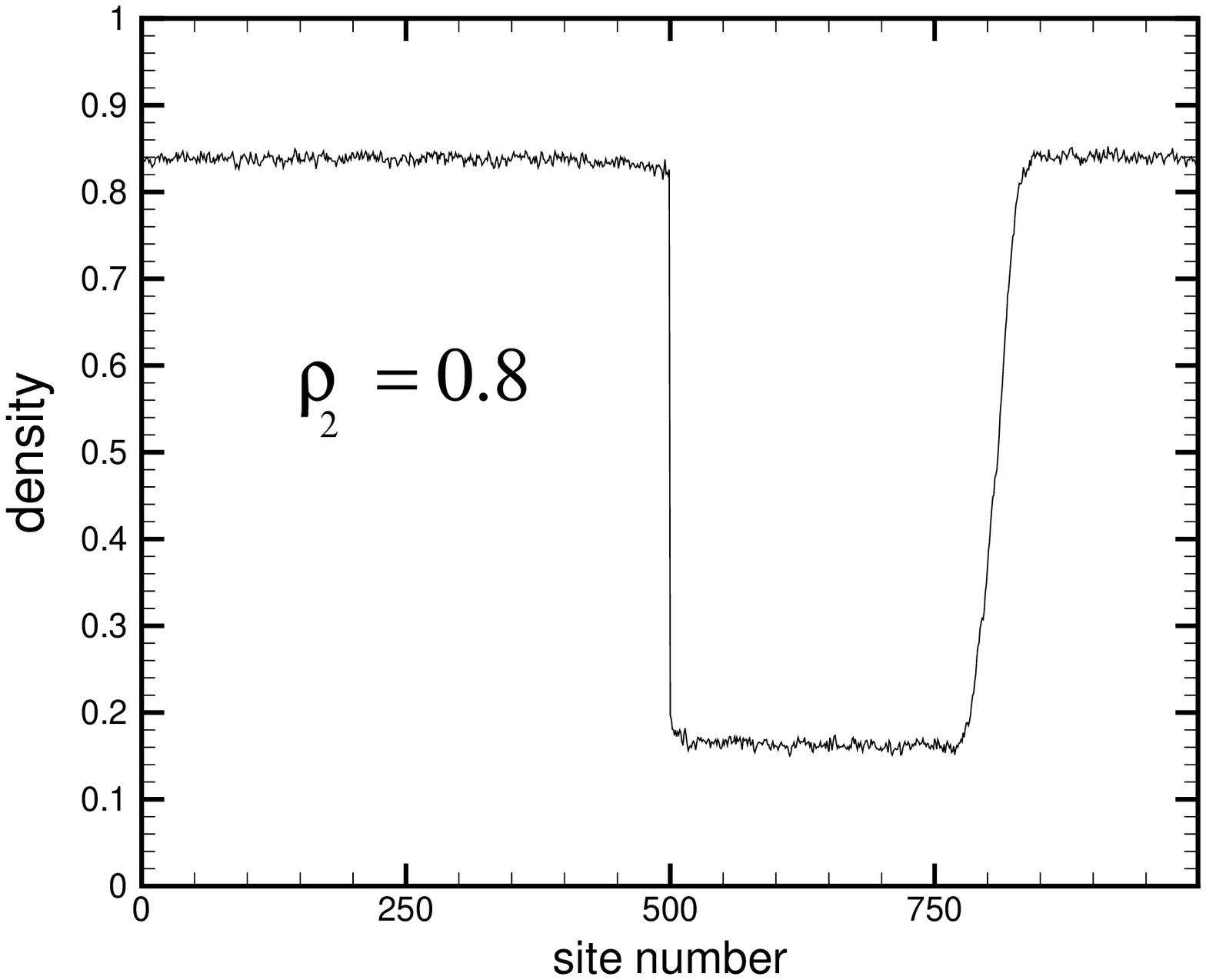}
\caption{Density profile by MC simulation in the first chain:
$\rho_1=0.2$ and $\rho_2=0.4$ (top) and $\rho_1=0.6$ and
$\rho_2=0.8$ (bottom). } \label{fig:bz2}
\end{figure}

In figure (4) top, the second chain affects the first chain's
density profile only on a local scale. The reason is that the
$\rho_1$ is below the limit to be globally affected. On the other
hand, if one increases $\rho_1$ to a higher value e.g. $0.6$ (figure
(4) bottom), it turns out that density profile of the first chain
gives rise to segregation between a high and a low density regions.
The formation of phase-segregated regime depends on the mutual
values of global densities in both chains. The above observations
have resemblance to the BML model of city traffic \cite{bml} in the
sense that below a given density, the conflicting flows do not
affect each other much and the cars are not notably blocked by the
other lane flow. Nevertheless, it should be mentioned that updating
rules and road structures in the BML model are entirely different to
our model's. Furthermore, the blocking mechanism in the BML model is
not only due to exclusion principle but is also related to the
cooperative motion of vehicles. In our model, the it is only the
exclusion at the crossing point which gives rise to blocking. To
support our density profiles findings, which have been obtained by
Monte Carlo simulations, we have numerically solved the mean-field
equation in the constant-density scheme described in section II.B.
The results are satisfactory and in qualitative agreement with Monte
Carlo simulations. As an example, in figure (5) we have sketched and
compared the profile of density in chain one for given global
densities in both chain both with Monte Carlo and numerical mean
field approach.

\begin{figure}
\centering
\includegraphics[width=7.5cm]{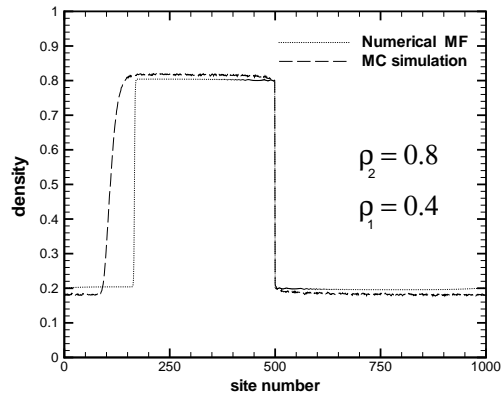}
\caption{Density profile in the first chain obtained via MC (dashed)
and MF (dotted). $\rho_1=0.4$ and $\rho_2=0.8$. } \label{fig:bz2}
\end{figure}

We observe that numerical mean field has qualitatively reproduced
the phase-segregated behaviour. The difference between high and
low densities given by MC and MF are in fairly well agreement.
However, there is a notable difference in the vicinity of the
domain wall. The prediction of MF is much sharper than that of
MC. The more smooth transition from low to high density in MC is
related to involve fluctuations which are not captured by the MF
approach especially in low dimensions. In figure (6) we show
density profiles in uniform phases.

\begin{figure}
\centering
\includegraphics[width=7.5cm]{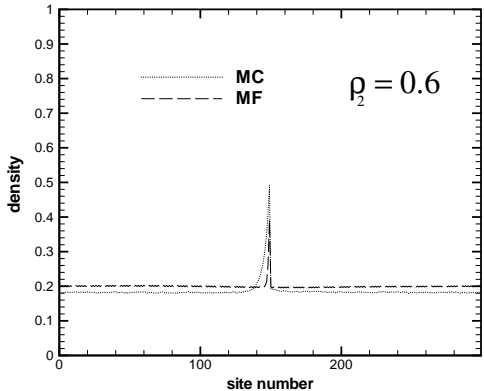}
\caption{Density profile in the first chain obtained via MC and MF.
$\rho_1=0.2$ and $\rho_2=0.6$. Number of iterations is $2000$. }
\label{fig:bz2}
\end{figure}

The predictions of MF and MC are reasonably close to each other.
Some comments on the constant-density approach seems unavoidable.
In each iterative scheme, the stability of long time behaviour
should be investigated. In our case, the choice of initial values
of density profiles play an important role. We took their values
within a small interval of size $\eta$ around the corresponding
global densities. In figure (6) we set $\eta=0.01$. We note that
the long time behaviour of the system of equations give rise to
an oscillatory pattern of profile. In numerical terms, the
iteration becomes unstable for large times. However, in
intermediate times, iterative method gives a reasonable answer.
In figure (7) we have depicted the behaviour of the density
profiles for a larger iteration. We have varied the the number of
iterations $M$ for $2000, 5000$ and $10000$ (from bottom to top)
correspondingly. For clarity, densities are shifted upwards for
$M=5000$ and $M=10000$.

\begin{figure}
\centering
\includegraphics[width=7.5cm]{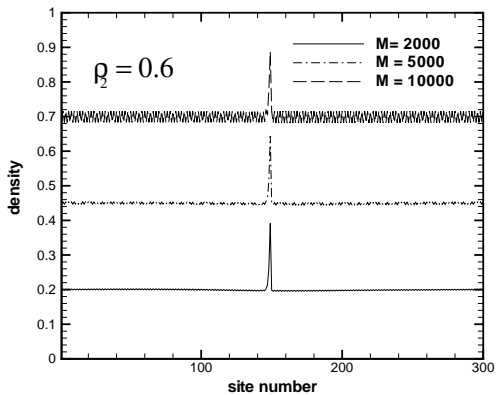}
\caption{Density profile in the first chain obtained via numerical
MF. $\rho_1=0.2$ and $\rho_2=0.6$. } \label{fig:bz2}
\end{figure}

We now turn into the general behaviour of density profiles. It
would be a natural question to ask under what circumstance
yielding leads to traffic jam formation in the first chain before
the crossing point. To this end, by Monte Carlo simulations we
have systematically surveyed the entire phase space
$\rho_1-\rho_2$ in the grids of $\delta \rho = 0.05 $ . Two
phases are identified: uniform (homogeneous) and
phase-segregated. In the uniform phase, the interaction between
two chains has only a local effect on profile of the first chain
whereas in the segregated phase, a domain wall separates a low
density and a high density region. Figure (8) exhibits the phase
structure of the problem. By symmetry, one obtains the phase
structure of the second chain via replacing $1 \leftrightarrow 2
$.

\begin{figure}
\centering
\includegraphics[width=7.5cm]{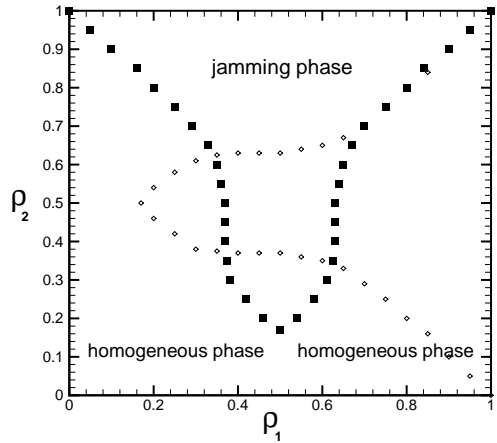}
\caption{Phase diagram of the model for the first chain. $\diamond$
shows the corresponding diagram for the second chain. }
\label{fig:bz2}
\end{figure}

\section{Mean-field phase diagram: Analytical approach }

Here we try to outline an analytical approach to obtain the model
phase diagram. This approach is based on the concept of density
phase segregation proposed in \cite{lebowitz1,lebowitz2}. Suppose
for given densities $\rho_1$ and $\rho_2$ one has density phase
segregation in both chains. We denote the high and low densities
in the first chain by by $\rho_h$ and $\rho_l$ and in the second
chain by $\xi_h$ and $\xi_l$ respectively. Also let us denote the
relative length of low and high density regions in the first
chain by $a_l$ and $a_h$ and in the second chain by $b_l$ and
$b_h$. We denote the probability of occupation of the crossing
site by a car from the first (second) chain by $r_1~(r_2$).
Therefore the probability that the crossing site will be empty is
$1- r_1 - r_2$. We can write the following mean field equations:

\be J_1=\rho_l(1-\rho_l);~~~~~J_2=\xi_l(1-\xi_l). \ee

\be J_1=\rho_h(1-\rho_h);~~~~~J_2=\xi_h(1-\xi_h) \ee

\be J_1=\rho_h(1-r_1-r_2);~~~~J_2=\xi_h(1-r_1-r_2) \ee

\be J_1=r_1(1-\rho_l);~~~~J_2=r_2(1-\xi_l) \ee

\be a_l+a_h=1;~~~~~b_l+b_h=1  \ee

\be \rho_l a_l + \rho_h a_h=\rho_1;~~~~\xi_l b_l + \xi_h
b_h=\rho_2 \ee

The above equations are easily solved and we find:

\begin{eqnarray}
r_1=r_2=\frac{1}{3};~\rho_l=\xi_l=\frac{1}{3};~\rho_h=\xi_h=\frac{2}{3};~J_1=J_2=\frac{2}{9}
\label{eqn12}
\end{eqnarray}

One also finds: $a_l=2-3\rho_1$ and $b_l=2-3\rho_2$. The
conditions $0\leq a_l \leq 1$ and $0\leq b_l \leq 1$ imply:
$\frac{1}{3} \leq \rho_1,\rho_2 \leq \frac{2}{3}$. In figure (9),
we have plotted the phase diagram. In the interior region, one has
density phase segregation in both streets. We note that middle
square region is in well qualitative agreement with the
prediction of MC simulations ( intersection of two curves in fig8
).

\begin{figure}
\centering
\includegraphics[width=7.5cm]{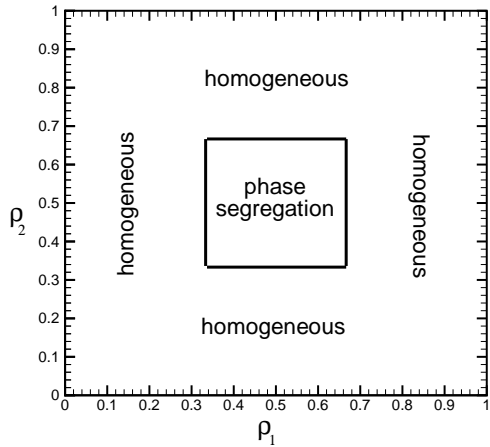}
\caption{Phase diagram of the model. } \label{fig:bz2}
\end{figure}

\section{Summary and Concluding Remarks}

We have investigated the characteristics of two conflicting
traffic flows within the framework of asymmetric simple exclusion
process. Two perpendicular chains interact with each other via
the intersection point. Using Monte Carlo simulations and
numerics, we have obtained the dependence of each chain current
on its own and on its perpendicular chain global density. It is
verified that the chains can maintain large currents up to rather
a high density. Interaction of two chains can effectively be
considered as a dynamic impurity. For some values of global
densities in the chains, the interaction of chains leads to
formation of high density region behind the intersection point
which is segregated from a low density region afterwards. By a
systematic scanning of the phase space, we have obtained the
structure of model phase diagram. Two phases of jamming (density
segregated) and regulated (uniform density) flows are identified.

\section{acknowledgement}

Fruitful discussions with Mustansir Barma, Gunter Sch\"{u}tz,
David Mukamel and Anatoly Kolomeisky are appreciated. Special
thanks are given to Majid Arabgol and Soltaan Abdol Hossein for
useful helps.

\bibliographystyle{unsrt}

\begin{thebibliography}{99}



\bibitem{zia} B. Schmittmann and R.K.P. Zia in: {\em Phase transitions and Crtitical Phenomena}, {\it vol 17}, ed.
C. Domb and L. Lebowitz (London: Academic) 1995.



\bibitem{schutz1} G. Sch\"{u}tz in: {\em Phase transitions and Crtitical Phenomena}, {\it vol 19}, ed.
C. Domb and L. Lebowitz, Academic Press, 2001.



\bibitem{kernerbook} B. Kerner, in {\em Physics of traffic flow}, Springer, 2004.



\bibitem{css99} D. Chowdhury, L. Santen and A. Schadschneider, {\em Physics Reports}, {\bf 329}, 199 (2000).



\bibitem{helbing} D. Helbing, {\em Rev. Mod. Phys.}, {\bf 73}, 1067 (2001).






\bibitem{bml} O.\ Biham, A.\ Middleton and D.\ Levine, Phys.Rev. {\bf A }{\bf 46}, R6124 (1992).



\bibitem{nagatani} T. Nagatani, {\em Phys. Rev. E} {\bf 48}, 3290 (1993).


\bibitem{cs} D.\ Chowdhury and A.\ Schadschneider, {\em Phys. Rev. E} {\bf 59}, R 1311 (1999).



\bibitem{foolad1} M.E. Fouladvand and M. Nematollahi, {\em Eur. Phys. J. B} {\bf 22}, 395 (2001).



\bibitem{foolad2} M.E. Fouladvand, Z. Sadjadi and M.R. Shaebani {\em J. Phys. A: Math. Gen}, {\bf 37}, 561 (2004).



\bibitem{foolad3} M.E. Fouladvand, Z. Sadjadi and M.R. Shaebani {\em Phys. Rev. E}, {\bf 70}, 046132 (2004).



\bibitem{ray} B. Ray and S.N. Bhattacharyya, {\em Phys. Rev. E}, {\bf 73}, 036101 (2006).



\bibitem{xiong} C. Rui-Xiong, Bai Ke- Zhao and L Mu-Ren, {\em Chinese Physics}, 15, No 7, July 2006.



\bibitem{gershenson} S-B Cools, C. Gershenson and B. D Hooghe, arXive: nlin.AO/0610040



\bibitem{foolad4} M.E. Foulaadvand and S. Belbaasi, {\em J. Phys. A: Math. Theor.}, {\bf 40}, 8289 (2007).



\bibitem{mcdonald} J.T. MacDonald, J.H. Gibbs and A.C. Pipkin {\em Biopolymers}, {\bf 6}, 1  (1968).



\bibitem{derrida1} B. Derrida, M.R. Evans, V. Hakim and V. Pasquir {\em J. Phys. A}, {\bf 26}, 1493 (1993).



\bibitem{derrida2} B. Derrida {\em Physics Report}, {\bf 301}, 65 (1998).



\bibitem{lebowitz1} S. Janowsky and J. Lebowitz {\em Phys. Rev. A}, {\bf 45}, 618 (1992).



\bibitem{lebowitz2} S. Janowsky and J. Lebowitz {\em J. Stat. Phys.}, {\bf 77}, 35 (1994).



\bibitem{kolomeisky1} A.B. Kolomeisky, {\em J. Phys. A: Math, Gen.}, {\bf 31}, 1153 (1998).



\bibitem{evans1} M.R. Evans, {\em J. Phys. A: Math,Gen.}, {\bf 30}, 5669 (1997).



\bibitem{chou} T. Chou and G. Lakatos, {\em Phys. Rev. Lett.}, {\bf 93}, 198101 (2004).



\bibitem{lakatos1} G. Lakatos, T. Chou and A.B. Kolomeisky, {\em Phys. Rev. E}, {\bf 71}, 011103 (2005).



\bibitem{foolad5} M.E. Foulaadvand, S. Chaaboki and M. Saalehi {\em Phys. Rev. E}, {\bf 75}, 011127 (2007).



\bibitem{pronina} E. Ekaterina and A. B. Kolomeisky, {\em J. Stat. Mech.}, {\bf 07}, P07010.  (2005).


\bibitem{barma1} G. Tripathy and M. Barma, {\em Phys. Rev. Lett.}, {\bf78}, 3039 (1997).



\bibitem{barma2} G. Tripathy and M. Barma, {\em Phys. Rev. E}, {\bf58}, 1911 (1998).



\bibitem{ns92} K. Nagel, M. Schreckenberg, {\em J. Phys. I France} {\bf 2}, 2221 (1992).



\bibitem{chung} K.H. Chung and P.M. Hui, {\em J. Phys. Soc. Jap.}, {\bf 63}, 4338 (1994).



\bibitem{yukawa} S. Yukawa, M. Kikuchi and S. Tadaki {\em J. Phys. Soc. Jap.}, {\bf 63}, 3609 (1994).



\end{thebibliography}

\end{document}